\documentclass[conference]{IEEEtran} 

\usepackage[T1]{fontenc} 
\usepackage{amsmath,amssymb,physics} 
\usepackage{algorithm}
\usepackage{algpseudocode}
\usepackage{float}
\usepackage{graphicx} 
\usepackage{cite} 
\usepackage{balance} 
\usepackage{qcircuit}
\usepackage{subcaption}
\usepackage{todonotes}
\usepackage{comment}
\usepackage{siunitx}
\usepackage{booktabs}
\usepackage{blindtext}
\usepackage{svg}
\usepackage{url}
\usepackage{xurl}

\usepackage{array}
\usepackage{balance} 
\usepackage{blindtext}
\usepackage{bm}
\usepackage{booktabs}
\usepackage{comment}
\usepackage{datetime}
\usepackage{fancyhdr}
\usepackage{float}
\usepackage{graphicx} 
\RequirePackage[colorlinks=true, allcolors=blue]{hyperref}
\usepackage{makecell}
\usepackage{multirow}
\usepackage{mwe}
\usepackage{siunitx}
\usepackage{subcaption}
\usepackage{svg}
\usepackage{tablefootnote}
\usepackage[absolute]{textpos}
\usepackage{url}
\usepackage{xcolor}
\usepackage{xurl}

\definecolor{ionqorange}{HTML}{FF5000}
\hypersetup{
    colorlinks=true,
    linkcolor=ionqorange,    
    citecolor=ionqorange,    
    urlcolor=ionqorange      
}

\IEEEoverridecommandlockouts 

\begin{document}

\title{Quantum Large Language Model Fine-Tuning}
\author{
    \IEEEauthorblockN{
           Sang Hyub Kim,
        Jonathan Mei,
        Claudio Girotto,
        Masako Yamada,
        Martin Roetteler
    }
    \IEEEauthorblockA{IonQ Inc., 4505 Campus Dr, College Park, MD 20740, USA}
}

\maketitle

\begin{abstract}

We introduce a hybrid quantum-classical deep learning architecture for large language model fine-tuning. The classical portion of the architecture is a sentence transformer that is powerful enough to display significant accuracy for complex tasks such as sentiment prediction. 
The quantum portion of the architecture consists of parameterized quantum circuits that utilize long-range connections between qubits.

We analyze the performance of the hybrid models for various settings of hyperparameters, including the number of qubits, the depth of the quantum circuits, learning rate, number of re-uploading steps, etc. 
Based on a screening study of main effects, we show an overall improvement in prediction accuracy over a comparable classical baseline, with a trend of increasing accuracy with number of qubits. We observe up to {$\mathbf{3.14\%}$} improvements in accuracy over classical architectures of comparable model size, within the set of hyperparameters probed in this study.

We demonstrate the contribution of each module in our architecture through ablation studies. Our studies are based on finite shot-counts and include simulations based on noisy quantum gates.\\[1ex]

\end{abstract}

\section{Introduction}

Large Language Model (LLM) fine-tuning is a process for adapting pre-trained models to specific tasks or domains. As a method for transfer learning, fine-tuning typically involves further training a pre-existing LLM on a smaller, task-specific dataset to enhance its performance and relevance. This process leverages the general knowledge already acquired by the LLM during its initial training on vast amounts of data.

We explore the integration of quantum computing into LLM fine-tuning. Hybrid quantum-classical approaches aim to leverage quantum circuits to enhance the model's classification capabilities, particularly in tasks involving complex data correlations. Specifically, we study replacing the classification head of a classical model with parameterized quantum circuit so that the new hybrid quantum/classical models are fine-tuned for sentiment analysis tasks. These quantum layers can potentially capture non-local correlations in data more effectively than classical networks, leading to improved accuracy, especially in low-data regimes. 

We hypothesize that the model's expressivity and thus accuracy can be enhanced by introducing a quantum machine learning (QML) layer, particularly when dealing with nonlocal data correlations. Case in point is the LLM for token prediction called the Sentence Transformer \cite{reimers_sentence-bert_2019}, for which \cite{tunstall_efficient_2022} produces a pre-trained model that we refer to as SetFit. To SetFit we add a final layer composed of a differentiable quantum circuit. The core idea is to use quantum systems that support operations beyond simply nearest-neighbor pairs of qubits \cite{chen_benchmarking_2024}.

The idea of fine-tuning a pre-trained neural network has evolved in tandem with the modern resurgence of neural networks \cite{jung_joint_2015,kading_fine-tuning_2017, wang_growing_2017}. In language processing, \cite{devlin_bert_2019} demonstrated that a single pre-trained BERT model has the ability to adapt to many different downstream tasks via fine-tuning with a strong showing on the GLUE benchmark \cite{wang_glue_2019}. Optimizing the pre-training process became its own field of interest \cite{sun_how_2019}. Even with the advent of powerful generalist ``few-shot'' pre-trained models \cite{brown_language_2020} that can seemingly handle disparate domains out of the box, there is ample evidence that fine-tuning can further improve model performance on unseen data \cite{liu_few-shot_2022,tunstall_efficient_2022}. This has implications for use cases requiring personalization where privacy matters and sensitive data is not otherwise widely available for pre-training. Examples include medicine where a model may be fine-tuned to encapsulate an individual patient's health history, chemistry where a model trained on molecular ground-state energies may be fine-tuned to predict other structural or functional properties, industrial applications where a company could find it useful to have a model that can reason about its own trade secrets, or even national security where a model is fine-tuned to understand secret classified intelligence material.

Our main contribution is a study of the accuracy of these hybrid quantum-classical models under different scaling regimes, under shot-noise and quantum gate-noise, and comparing against multiple classical methods such as Support Vector Classifiers (SVC), logistic regression, and perceptrons. Our experiments, conducted with noisy state vector simulators \cite{wang_quantumnat_2022}, indicate that certain configurations of the QML head achieve higher accuracy than the classical methods for sentiment analysis benchmarks like the Stanford Sentiment Treebank (SST2) \cite{socher_recursive_2013}.

We believe that this research represents a step towards understanding the potential for quantum computing to enhance AI capabilities for complex real-world problems.

\section{Preliminaries}
We first cover the requisite background for our method. This includes some concepts from both classical machine learning as well as QML.

\subsection{Sentence transformer and SetFit}

In recent years, frontier causal (or autoregressive) language models have gained significant attention for their ability to form complex, coherent sentences and competently perform various tasks across multiple disciplines. These tasks may appear unrelated to usage of language, apart from the availability of pedagogical material describing them in plain natural language on the internet.
From a technical perspective, transformers \cite{vaswani_attention_2017} form the basis of these modern language models \cite{brown_language_2020}. 
Currently, the enormous amount of resources required to develop such large transformer-based models from scratch has resulted in more decentralized effort focused on engineering the systems around these LLMs, a significantly more accessible task.

One such example can be seen in the paradigm of retrieval augmented generation (RAG) \cite{guu_retrieval_2020, lewis_retrieval-augmented_2020}, 
a framework where an external knowledge base, separate from the training data, informs the output generated by the LLM.
The strategy enables improved interpretability and alignment with a curated source of factual data, reducing the so-called ``hallucination'' problem, and enhanced comprehension of data that is private or specialized.
Often, the knowledge base is textual and is converted to embedding vectors before being consumed by the LLM. It is clear then that models that create the embeddings, often non-causal rather than autoregressive, are as crucial to the system as the LLM itself.

One of the most well-known transformer-based embedding models is BERT \cite{devlin_bert_2019}, which has been used successfully in RAG environments and serves as the foundation for numerous derived models. Among the different encoder-based architecture derived from BERT is the Sentence Transformer \cite{reimers_sentence-bert_2019}. By leveraging siamese and triplet networks, the Sentence Transformer produces embeddings of sentences that behave in more semantically meaningful ways than previous embedding methods\cite{pennington_glove_2014}.

SetFit \cite{tunstall_efficient_2022} introduces a few-shot fine-tuning approach for the Sentence Transformer based on contrastive learning. This technique is particularly useful in low-data regimes, which are of particular interest in QML due to practical data throughput limitation of the current quantum hardware. Typical training set sizes are 128, 256, or 512 samples, typical test set sizes are 8,192 samples. The SetFit token prediction base model has about 110 million parameters. We build on this base model and fine-tune it based on training sets of size 256 and test sets of size 8,192.  

\subsection{Quantum computing}

A quantum computation, specified as a circuit, is a series of logic gates and measurements performed on a qubit register. All modern quantum computers implement an entangling gate. A universal gate set for quantum computing consists of a single 2-qubit entangling gate and a small core set of single-qubit gates \cite{nielsen_quantum_2002}.
Entanglement, a critical feature of QC, distinguishes it from classical computing: characterizing the probability distributions from measuring quantum circuits with high degrees of entanglement can be exponentially hard for classical computers.

The underlying configuration of qubits at any point in a circuit can be mathematically modeled as a vector known as the state vector, assuming ideal, noiseless operation, while gates can be described as unitary matrices that act on these vectors. For example, the gate representing a rotation about the $y$-axis can be written 
\begin{align*}
    R_Y(\theta)=\begin{bmatrix}
        \cos(\theta/2) & -\sin(\theta/2) \\ \sin(\theta/2) & \phantom{-}\cos(\theta/2)
    \end{bmatrix}.
\end{align*}
Since products of unitary matrices are also unitary, multi-qubit gates (involving more than 2 qubits) can be defined via compositions of 1-qubit and 2-qubit gates. Therefore, any arbitrary gates or groups of gates can be represented by the letter $U$, denoting unitary.

\subsection{Quantum machine learning}

Early work in QML addressed traditional ML tasks often with custom algorithms for each task \cite{guta_quantum_2010, schuld_quantum_2014-1}. The advent of optimization-based approximation methods in chemistry such as the variational eigensolver \cite{peruzzo_variational_2014} and analytic techniques for evaluating gradients on gate-based quantum computers \cite{li_hybrid_2017} connected these quantum methods to the nascent differentiable revolution happening in the classical realm \cite{krizhevsky_imagenet_2012}.

In this context, we do not attempt to provide comprehensive coverage of the QML field, and rather emphasize the key features that relate to contemporary machine learning and the challenges that  are pertinent to our investigation.

Classically, there is a substantial body of work that relate the learning and generalization of deep neural networks to kernel machines \cite{lee_wide_2019, atanasov_neural_2021}.
Recent research demonstrated that a large class of QML modules also exhibit an equivalence with classical kernel methods \cite{schuld_supervised_2021}. Furthermore, it has been empirically observed that optimally tuned QML models using specific data encodings converge to well-establsihed classical kernels \cite{ablan_similarity_2025}. Consequently, the data encoding and ansatz used for the QML module are of central importance in attaining behavior and, therefore, performance that surpasses what is classically achievable.

\begin{figure*}[hbt]
    \centering\includegraphics[width=1\linewidth]{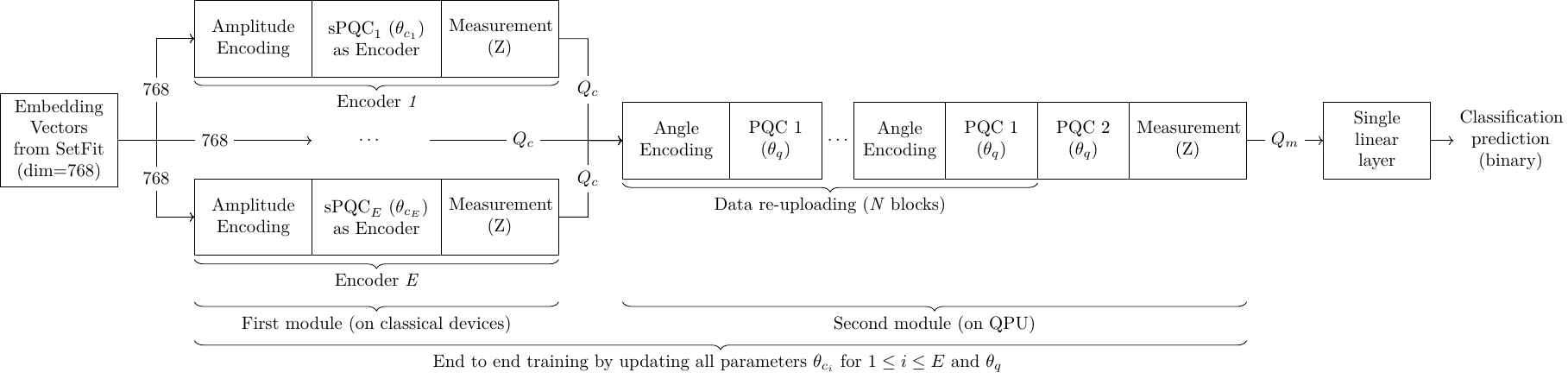}   
    \caption{Block diagram of the last layer added to the base LLM, consisting of two modules. The first module, implemented on classical devices, contains a variable number of $E$ parallel encoders which encode the embedding vectors from SetFit of dimension equal to $768$ into output vectors of dimension $Q_c$. In the case of a single encoder $E = 1$. The second module, implemented on QPU, contains a flexible data re-uploading module with a variable number $N$ of repetitions. The output of the end-to-end trainable module is a vector with dimension $Q_m$, the number of qubits measured, where $Q_m$ is fixed at 1 for all experiments.}
    \label{fig:full_block_diagram}
\end{figure*}

Additionally, there is an interesting branch of QML that can be collected under the term ``quantum-inspired methods'', which takes concepts and computational tools from quantum mechanics and computing. The methods are applied classically to traditional ML problems ranging from (but not limited to) commutators for optimization of algebraic signal processing operators \cite{mei_signal_2017} to tensor networks for image classification \cite{stoudenmire_supervised_2016}. Our architecture, fully described in Figure~\ref{fig:full_block_diagram}, leverages both QML and quantum-inspired methods.

\subsubsection{Angle encoding}
\label{sec:tech_prelim:qml:angle_encoding}
We achieve a high degree of expressivity in our quantum data representation by using angular encoding \cite{schuld_supervised_2021}. In this scheme, classical data determines the phases of qubits. The qubits are rotated around the Bloch sphere by an angle proportional to the respective data values
\[
\mathbf{x}=\begin{bmatrix}x_1 & x_2 & \ldots & x_n\end{bmatrix}^\top.
\]
An example of angular encoding, using the y-axis is:
\[
\ket{\psi(\mathbf{x})} = R_Y(x_1) \otimes R_Y(x_2) \otimes \dots \otimes R_Y(x_n) \ket{0}^{\otimes n}.
\]
where we overload $\otimes$ as the outer product when used as a binary operator and to denote an iterated outer product with itself when used in the exponent.

\subsubsection{Quantum data re-uploading}
To increase the expressivity of angular encoding, we combine it with re-uploading \cite{perez-salinas_data_2020}. This is a meta-scheme where the chosen encoding is applied both at the beginning of the quantum circuit and also repeated and possibly interspersed with other computational parts of the circuit.

\subsubsection{Ansatz}
\begin{figure}[hbt]
    \begin{subfigure}[c]{0.7\linewidth}
    \centering
        \includegraphics[width=\linewidth]{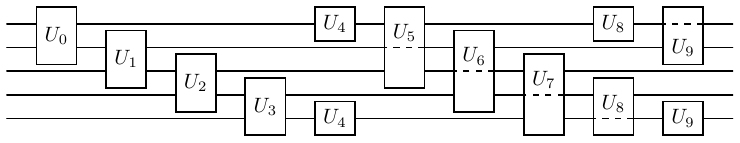}
    \end{subfigure}
    \hfill 
    \begin{subfigure}[c]{0.2\linewidth}
    \centering
        \includegraphics[width=\linewidth]{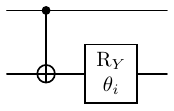}
    \end{subfigure}
    \caption{The ansatz (left) consists of layers with increasing connectivity $C$ across the width of the circuit. Each block $U_i$ is defined as a combination of a controlled NOT and a single-qubit rotation $R_Y$ through angle $\theta_i$ around the Y-axis (right).}
    \label{fig:unitary}
\end{figure}

Figure \ref{fig:unitary} shows snippets of the ansatz used as a fundamental building block in the quantum circuit, which consists of layers with varying connectivity spanning the width of the circuit. This design facilitates operation on hardware platforms that can operate beyond nearest neighbor pairs of qubits \cite{chen_benchmarking_2024}. Furthermore, as the number of qubits (the circuit \textit{width}) increases , classical simulation of the circuit becomes increasingly difficult. 

\begin{figure*}[hbt]
    \centering\includegraphics[width=0.9\linewidth]{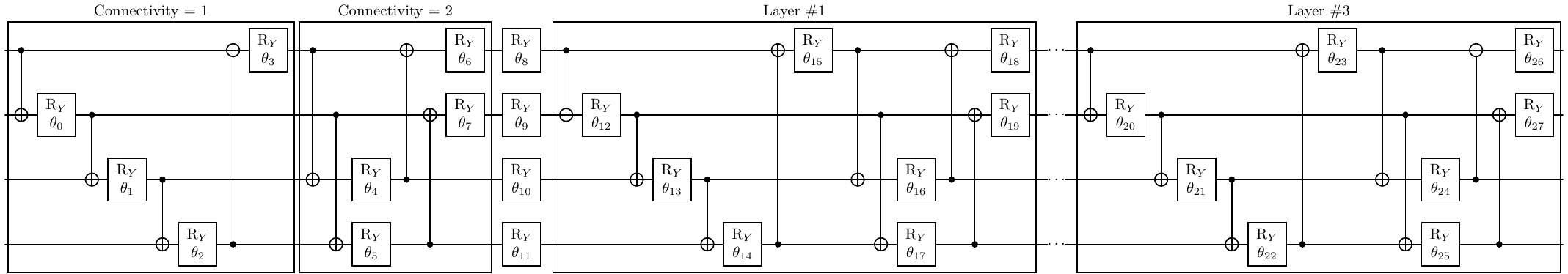}    
    \caption{Example of a PQC with $Q=4$ qubits, 3 layers, and connectivity $C=2$.}
    \label{fig:ex_pqc}
\end{figure*}

The complete solution can be specified by a larger parameterized quantum circuit, as shown in Figure \ref{fig:ex_pqc}.

\section{Methodology}

The goal of our investigation is to fine-tune a SetFit model using a quantum module as a classification head, whose novelty we describe in details in the following sections. 

\begin{figure}[hbt]
    \centering\includegraphics[width=0.8\linewidth]{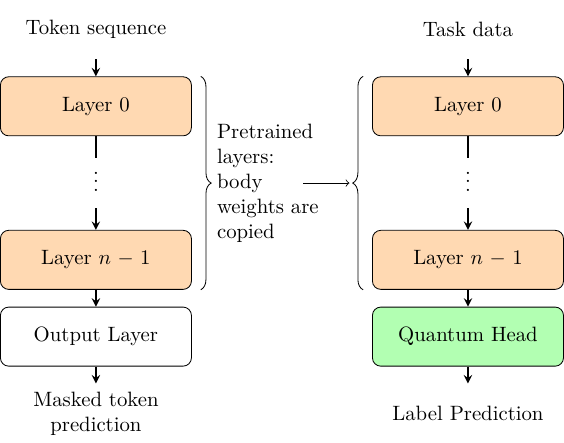}
    \caption{The overall Architecture, which includes both embedding and transformer blocks, in orange, and is composed by the pretrained classical masked language model (left), and the quantum classification head (right).}   
    \label{fig:overall_arch}
\end{figure}

\subsection{Overall Architecture}
The architecture in Figure~\ref{fig:overall_arch} shows at a high level how the fine-tuning procedure incorporates the quantum head as the last layer on top of the classical layers that inherit the weights from the pre-trained model. It should be noted that in this study, we did not re-train the weights of the lower, i.e., we kept the weights of the base model frozen. Re-training the LLM is a possibility for future research and might provide a path to even higher accuracy.

\subsection{Quantum-inspired latent vectors}

The hybrid classical-quantum architecture, used as the classification head for the fine-tuning task, combines classical processing with a Quantum Processing Unit (QPU) that performs dimensionality reduction and classification in an end-to-end training framework.

The process begins with high-dimensional embedding vectors ($\mathbb{R}^{768}$), in this case output by SetFit, that are initially processed by the first, classical module. This module employs a simulated amplitude encoding module and a simulated Parametrized Quantum Circuit (sPQC) with parameters $\theta_c$, where the subscript $c$ denotes correspondence to a classical device. A crucial step here is that the measurement output of the sPQC, specifically the expectation values of the observables (Pauli-Z) are computed exactly and used as a latent representation. This provides dimensionality reduction, similar to that of an autoencoder \cite{ranzato_sparse_2007}. Hence, we also refer to this component as an encoder.

The resulting latent vector, with a dimensionality corresponding to the number of classical qubits ($Q_C$) from sPQC, is then passed to the second module, which is intended to run on an actual QPU. This second module loads the latent vector via angle encoding, where information is encoded through single qubit rotations, as described in Section \ref{sec:tech_prelim:qml:angle_encoding}. The module consists of a second Parametrized Quantum Circuit (PQC) with parameters $\theta_q$, where the subscript $q$ denotes correspondence to a quantum device. A measurement (Pauli-Z) is performed on the output of PQC.

The measurement output of the second module is then passed through a single Linear layer. This linear layer has $(Q_C + 1) \times k$ parameters, and it generates the prediction logits for a $k$-class classification task.

Both the first and second modules are differentiable, meaning the parameters $\theta_c$ and $\theta_q$ can be optimized simultaneously in an end-to-end fashion to enable the model to learn dimensionality reduction in the first module and perform the classification task in the second module.

\subsection{Single-encoder head}

Inside the simulated quantum encoder block (sQE), we use  amplitude encoding to represent the classical feature vector passed from the last layer of the LLM. The classical data values are embedded into the amplitudes of the quantum state. Specifically, a classical vector $\mathbf{x}$ is encoded as:
\[
|\psi(\mathbf{x})\rangle = \sum_{i=1}^{n} x_i \ket{i}.
\]

Since the process is simulated, the state can be directly ``loaded'' without explicitly constructing a quantum circuit to prepare $|\psi(\mathbf{x})\rangle$. 

\subsection{Multi-encoder head}

We introduce a multi-encoder design to study the scaling behavior of sPQC as the number of qubits in the PQC increases. This approach allows for a more manageable increase in sPQC complexity as the PQC grows. Remember that sPQC, as a quantum-inspired method, is implemented via noiseless state vector simulation.

The implementation's memory complexity demonstrates the advantages of this solution. Using a single encoder with an $EQ$ qubit PQC results in $O(2^{EQ})$ memory complexity. By employing $E$ encoders, each with $Q$ qubits, the memory complexity of the multi-encoder sPQC becomes $O(E2^Q)$, offering a more favorable scaling than single-encoder heads.

\subsection{Energy consumption estimates}
This section estimates energy consumption for \textit{inference} of the second module (Figure~\ref{fig:full_block_diagram}) for both QPU and GPU, where GPU inference uses statevector simulation. As the first module is simulated on classical devices in both cases, the comparison of the second module shows how the difference in energy consumption can be expected to scale with qubit number for both QPU and GPU for the proposed quantum fine-tuning method. We use the same ansatz configurations which we used for qubit scaling experiments. For the GPU analysis, we use the published technical specifications for the NVIDIA H100 SXM \cite{noauthor_nvidia_nodate}. Other key details are explained in the Supplementary Information ~\ref{sec:energy_consumption_cmp}.

\begin{figure}[hbt]
    \centering\includegraphics[width=0.45\textwidth]{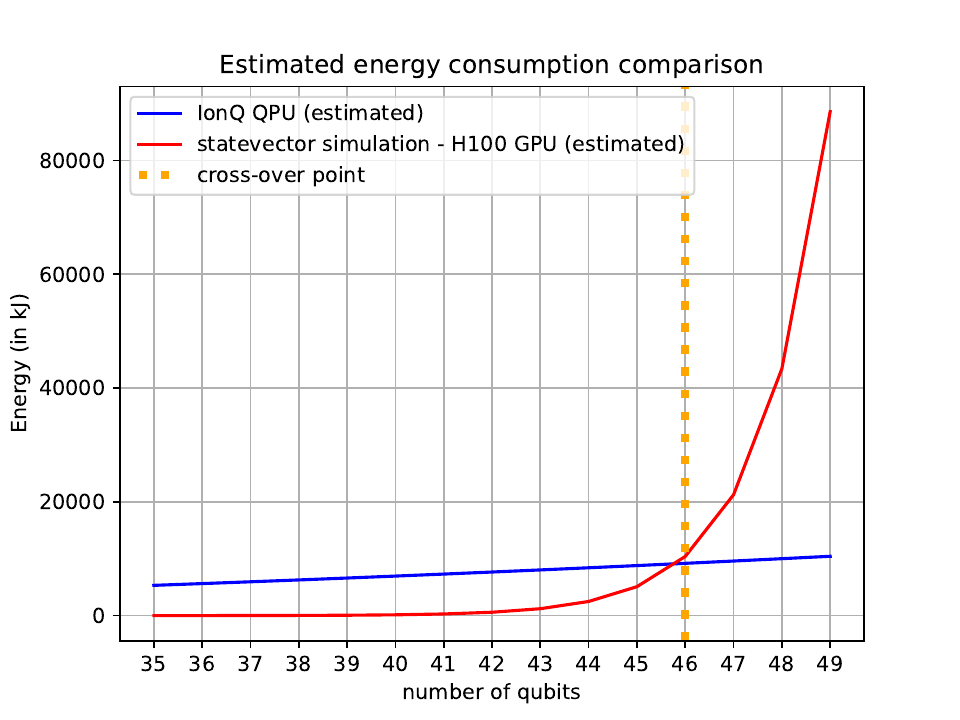}
    \caption{Estimated energy consumption for inference on GPU increases significantly faster than on QPU where GPU inference uses statevector simulation. The cross-over point occurs at 46 qubits. The ansatz configurations chosen are:
    \small
    $R=4$,
    $M=2$,
    $N=1$,
    }   
    \label{fig:energy_consumption_estimate}
\end{figure}

Figure~\ref{fig:energy_consumption_estimate} shows how energy consumption curves for inference scales with qubit number for both QPU and GPU. We find that inference on GPU increases significantly faster than on QPU with a cross-over at 46 qubits.

\subsection{Hyperparameters}
The model family described in the previous sections has multiple hyperparameters. A summary is provided in Table~\ref{tab:hyperparam_letters}. 

\begin{table}[hbt]
    \centering
    \begin{tabular}{c c c}
\specialrule{.1em}{.05em}{.05em} 
        \textbf{Hyperparameter} & \textbf{Symbol} & \\ [0.5ex] 
\specialrule{.05em}{.05em}{.05em}
        \underline{Q}ubits & $Q$ \\       
        Number of \underline{E}ncoders & $E$ \\         
        \underline{R}e-upload number & $R$ \\ 
        Number of \underline{M}ain blocks & $M$   \\ 
        \underline{N}umber of re-uploading blocks & $N$ \\    
        \underline{C}onnectivity & $C$\\ 
        \underline{B}atch Size & $B$ \\ 
        Number of \underline{S}hots & $S$ \\ 
        \underline{F}inal linear layer & $F$ \\
        \underline{L}earning rate & $L$ \\ 
        Learning rate decay & $\gamma$ \\ 
        Weight decay & $\rho$ \\ [1ex] 
\specialrule{.1em}{.05em}{.05em} 
    \end{tabular}
    \caption{Different hyperparamaters used in this paper.}
    \label{tab:hyperparam_letters}
\end{table}

\section{Experiments}

Our two primary goals are related. First, we aim to compare the performance of our model against a strong analogous classical baseline. Second, we want to study how our architecture behaves as we scale up the available quantum computational resources. As we work towards these goals, we also naturally validate the usefulness of different architectural components.

\subsection{Models, datasets, and tasks}

Following the SetFit case study \cite{tunstall_efficient_2022}, we use the Sentence Transformer pre-trained on \texttt{paraphrase} data\footnote{\url{https://huggingface.co/sentence-transformers/paraphrase-mpnet-base-v2}}.

We examine the Stanford Sentiment Treebank (SST2) \cite{socher_recursive_2013} dataset for a 2-label classification task, the version trimmed for SetFit (Supplementary Information ~\ref{sec:sst2_for_setfit}). We use custom splits of the dataset, similar to SetFit \cite{tunstall_efficient_2022}, operating in a low-data regime. We take $256$ positive and $256$ negative samples from the dataset that we split into $85\%$ for the ``training'' set and $15\%$ for the ``validation'' set.  We select models based on the accuracy evaluated on the validation set, then we plot and report accuracies evaluated on the remaining ``test'' split of the data set.

\subsection{Scaling}
We study the scaling behavior in relation to the number of qubits, number of main block layers, re-upload number, number of re-uploading layers.

\subsection{Ablations}
We perform ablation studies to quantify the individual contributions of each component to the observed benefits. First, we replace the sPQC layer with a neural network-based encoder, and vary its configurations and hyperparameters to evaluate the impact of the simulated quantum encoder on accuracy.
Second, we substitute the quantum fine-tuning based classification head with classical MLPs, sweeping through configurations and hyperparameters to assess whether quantum fine-tuning enhances the performance compared to a neural network-based classification head.
Finally, we remove the final linear layer and use the expectation value of a qubit in the z-basis for prediction.

\subsection{Noise}
Leveraging the paradigm of noise-aware training \cite{wang_quantumnat_2022}, we run noisy simulations with depolarizing gate noise and shot noise during both training and inference, with gate noise error rates derived from near-term quantum hardware \cite{noauthor_ionq_nodate}. This allows us to verify the applicability of our method on near-term real quantum device, demonstrating robust learning and inference in realistic operating conditions. TorchQuantum \cite{wang_quantumnat_2022} enables convenient simulations of the circuits with simulated noise while utilizing the automatic differentiation provided by PyTorch \cite{paszke_automatic_2017}.

Incorporating the effects of shot noise on state 
vector simulation involves sampling from a multinomial distribution. If the state vector being measured is given as
$|\psi\rangle = \sum_i a_i |i\rangle$,
then letting $p_i=|a_i|^2$, we have the distribution of $n$ shots following
\begin{align}
    \mathbf{y} \sim \textrm{Multinomial}(n, \mathbf{p}).
\end{align}

\begin{figure}[ht]
    \centering\includegraphics[width=.9\linewidth]{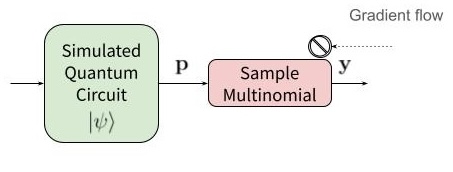}
    \caption{Straightforward sampler}
    \label{fig:shot_noise_blocked}
\end{figure}

However, we note that in the context of machine learning, sampling naively is not a differentiable operation with respect to its arguments (i.e. the probability vector $\mathbf{p}$) and blocks the gradient from reaching the parameterized circuit during backpropagation, as shown in Fig.~\ref{fig:shot_noise_blocked}.

\begin{figure}[ht]
    \centering\includegraphics[width=.7\linewidth]{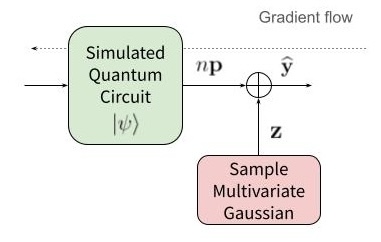}
    \caption{Differentiable sampler}
    \label{fig:shot_noise_differentiable}
\end{figure}

Instead, we implement an approximate sampler using the reparameterization trick \cite{kingma_auto-encoding_2014}. To apply this to quantum simulation, we take
\begin{align}
    \widehat{\mathbf{y}}-n\mathbf{p} \sim \mathcal{N}(0, \boldsymbol{\Sigma})
\end{align}
with $\boldsymbol{\Sigma}$ such that $\mathbf{y}\overset{d}{\rightarrow}\widehat{\mathbf{y}}$ as $n\rightarrow\infty$. This is shown in Fig. \ref{fig:shot_noise_differentiable}.
We can see that this reparameterization moves the non-differentiable sampling module off the path to the parameterized circuit, thus allowing gradients to flow through. As a post-processing step, we can rectify any nonnegative values to make the result physically consistent.

\subsection{Compute setup}
We performed experiments using 3 NVIDIA GPU types: L4 (24GB), A100 (80GB), and H100 (80GB). The H100s were used to perform the 18-qubit experiments for faster run times. All other experiments employed a combination of L4 and A100. 

We trained the quantum head for 800 epochs. Training the 10-qubit single sQE architecture took approximately 6.8 hours on an A100 or 9.1 hours on an L4. Training the 16-qubit single sQE architecture took 43 hours on an A100 or 56 hours on an L4. Training the 18-qubit single sQE architecture took 90 hours on an H100.

\section{Results}
We present the findings from our different experiments. An example of the training loss curve for one typical run is shown in Sec.~\ref{sec:training_loss}.
\subsection{Scaling studies}
\subsubsection{Accuracy over Qubits}

\begin{figure*}[hbt]
    \centering
    \begin{subfigure}[t]{0.45\textwidth}
        \centering
        \includegraphics[width=\textwidth]{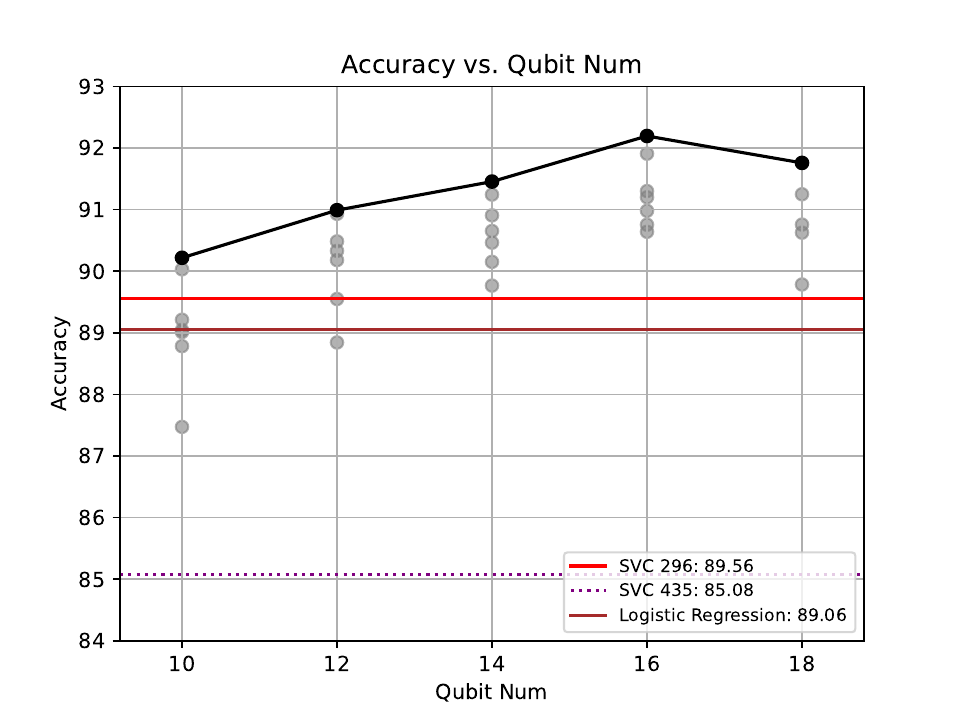}
        \caption{Single simulated quantum encoder: measured accuracy for hyperparameters
        \small
        $E=1$,
        $R=4$,
        $M=2$,
        $N=1$,
        $B=16$,
        $S=8192$,
        $\gamma=1.0$,
        $\rho=0.0$,
        $L\in\{1,\,1.5,\,2.5,\,3,\,5\}\times10^{-3}$.
        }
        \label{fig:acc_over_qubits_se}
    \end{subfigure}\hfill
    \begin{subfigure}[t]{0.45\textwidth}
        \centering
        \includegraphics[width=\textwidth]{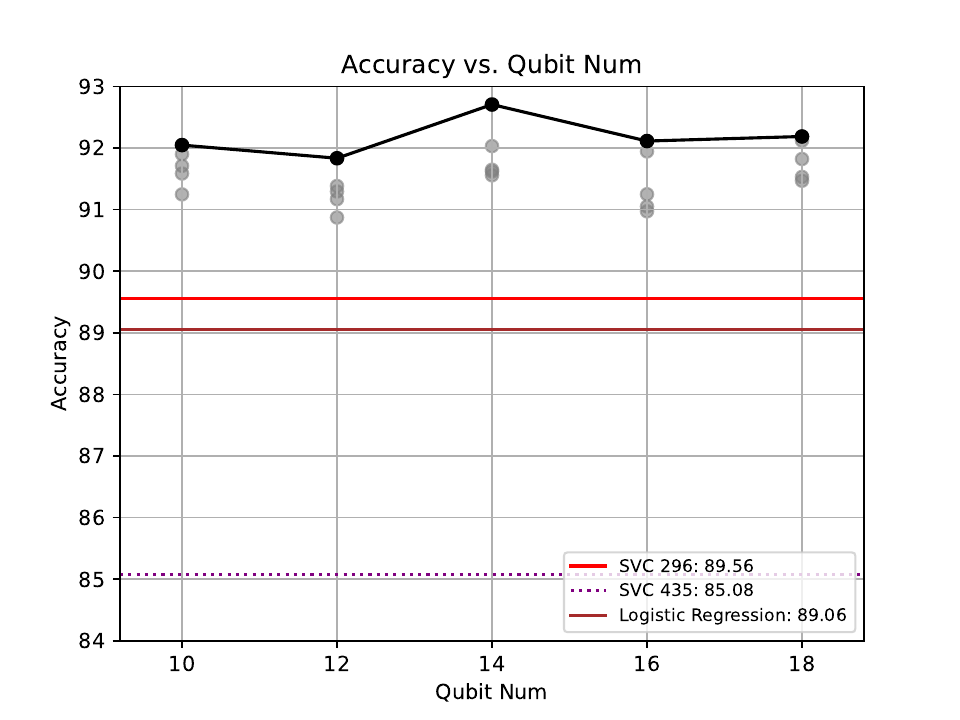}
        \caption{Multi simulated quantum encoder: measured accuracy for hyperparameters 
        \small
        $E=2$,
        $R=4$,
        $M=2$,
        $N=1$,
        $B=16$,
        $\gamma=1.0$,
        $\rho=0.0$,
        $S=8192$,
        $L\in\{1,\,1.5,\,2.5,\,3,\,5\}\times10^{-3}$.
        }   
        \label{fig:acc_over_qubits_me}
    \end{subfigure}
    \caption{Accuracy over the number of qubits. A black dot for each qubit number is the best accuracy from the sweep. The gray dots are the results from different hyperparameter settings for the learning rate $L$.   
        }    
    \label{fig:acc_over_qubits}
\end{figure*}

Figure~\ref{fig:acc_over_qubits_se} shows the trend of accuracy over the number of qubits for the single-encoder architecture. The gray dots represent results from different learning rates, with the black dots indicating the highest accuracy obtained at each number of qubits. Given finite resources, this qubit number sweep was conducted while varying the learning rate only and keeping the re-upload number, number of main block layers, and number of re-uploading layers fixed. Table~\ref{tab:config_se_sweep} provides a summary of the hyperparameters used for this training.

Figure~\ref{fig:acc_over_qubits_me} shows accuracy over the number of qubits for a multi-encoder architecture with two encoders. Similarly to the single encoder study, we varied the learning rate while keeping the re-upload number, number of main block layers, and number of re-uploading layers constant, as summarized in Table~\ref{tab:config_me_sweep}.

In the single-encoder architecture, accuracy shows an upward trend as the number of qubits increases, both in the encoder and the quantum data re-uploading block. It is worth noting that the observed accuracy at 18 qubits goes against the this trend, an artifact we attribute to fixing the particular re-upload number, number of main block layers, and number of re-uploading layers. Elsewhere in this study, we have observed that more extensive sampling of hyperparameters can yield gains in maximum accuracy by 0.5\% or more. We hypothesize that by floating all of the hyperparameters, the entire solid line (not exclusive to qubit 18) has margin to improve. 

The multi-encoder architecture, based on two encoders, consistently achieves higher accuracy scores than the single-encoder architecture, even at its peak accuracy at 16 qubits.
All configurations of the multi-encoder architecture reached the highest accuracy scores obtained in the study. This difference in behavior observed between the single-encoder and multi-encoder settings suggests that the encoders capacities in this scaling regime are being determined by their parameter counts rather than the sizes of their Hilbert spaces, as a multi-encoder architecture uses more parameters but has smaller Hilbert spaces than its analogous single-encoder architecture.
These saturated scores for the multi-encoder architecture could be limited by frozen LLM weights, LLM capacity, the complexity of the dataset, and encoder power. 

These results point to the need for larger simulation studies with more thorough explorations of relevant hyperparameters -- including number of encoders -- to examine how classification accuracy scales with number of qubits.

\subsubsection{Accuracy over re-upload number}
\begin{figure}[htb]
    \centering\includegraphics[width=0.45\textwidth]{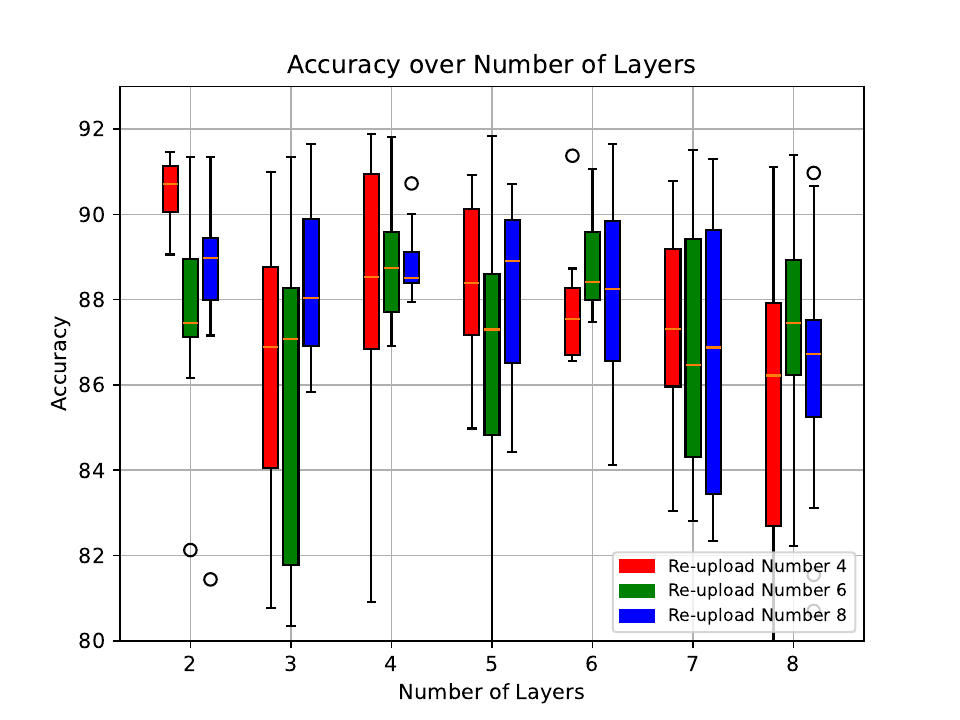}
    \caption{Multiple re-uploading steps and multiple layers have no discernible impact on accuracy. Box plots show the median (line), inter-quartile range spanning from the 1st to the 3rd quartile (box), the full range excluding outliers (whiskers), and outliers (dots). Hyperparameters chosen were:
        \small
        $Q=14$,
        $E=1$,
        $N=1$,
        $L\in\{2,\,2.5,\,5\}\times10^{-3}$,
        $B=16$,
        $\gamma=1.0$,
        $\rho=0.0$,
        $S=8192$.
        }
    \label{fig:acc_over_layer_num_main}
\end{figure}

In this section, we investigate the effect of re-upload number, an ansatz configuration parameter, on the accuracy trend. re-upload number refers to the number of repeats of data re-uploading, shown in Figure~\ref{fig:full_block_diagram}.
Fig.~\ref{fig:acc_over_layer_num_main} shows accuracy over number of main block layers across multiple re-upload numbers. We keep other hyperparameter for the ansatz configuration constant such as number of re-uploading layers. 
We compare multiple re-upload numbers and various number of main block layers across 3 learning rates. The configurations are summarized in Table~\ref{tab:config_hyper_cube_sweep_reupload_num}. The data points in the plot represent the best results from each configuration. 

Fig.~\ref{fig:acc_over_layer_num_main} does not reveal a discernible impact on accuracy coming from multiple re-uploading steps and multiple layers. When considered in conjunction with the results from Fig.~\ref{fig:acc_over_layer_num_reupload}, we observe that increasing re-upload number, number of main block layers, and number of re-uploading layers does not directly enhance accuracy. Notably, configurations with small Number of Layers and re-upload number remain still effective, and the smallest number of main block layers yields higher accuracy than larger ones. 

From these results and the ones in Figure~\ref{fig:acc_over_qubits_se}, which was obtained with fixed re-upload number, number of main block layers, and number of re-uploading layers, we deduce that increasing the number of qubits is a more effective strategy to improve the accuracy of the model.

\begin{table}[hbt]
  \centering
    \begin{tabular}{l @{\qquad} c @{\qquad} r} 
    \specialrule{.1em}{.05em}{.05em}  
    \textbf{Model} & \textbf{Maximum} & \textbf{Number of} \\
    \textbf{ } & \textbf{Accuracy} & \textbf{parameters} \\[0.5ex] 
    \specialrule{.05em}{.05em}{.5em} 
     Single sQE 10Q & 90.21 & 353 \\ 
     Single sQE 12Q & 90.99 & 423 \\ 
     Single sQE 18Q & 91.75 & 633 \\
     Multi sQE 10Q & 92.04 & 858 \\
     Multi sQE 14Q & \textbf{92.70} & 886 \\
     Multi sQE 18Q & 92.18 & 914 \\ 
     \hline 
     SVC 296 & 89.56 & 296 \\
     SVC 435 & 85.08 & 435 \\
     Logistic Regression & 89.06 & 769 \\ 
       Neural Network (0/0) & 91.05 & 1,538 \\ 
       Neural Network (1/48) & 91.22 & 37,010 \\ 
       Neural Network (1/96) & 91.36 & 74,018 \\ 
       Neural Network (1/144) & 91.36 & 111,026 \\
       Neural Network (1/192) & 91.44 & 148,034 \\ [1ex] 
    \specialrule{.1em}{.05em}{.05em}
    \end{tabular}
  \caption{Our hybrid quantum-classical models, as a function of qubit number, are compared against a selection of classical models, including Support Vector Classifiers (SVC)~\cite{svc}, logistic regression, and an ablation study including neural network with varying number of hidden layers and dimensions, indicated in parentheses.  Across all models, the highest accuracy is achieved with a multiple quantum encoder at 14 qubits, indicated in bold face. The best quantum model Multi sQE 14Q, achieves an improvement of $3.14\%$ over the best classical model considered in this study SVC 296 which has comparable model size\tablefootnote{We take the number of active support vectors as the effective number of parameters for the SVC using the \textit{dual} formulation, keeping in mind that by using the RBF kernel, the \textit{primal} formulation is infinite-dimensional, and so parameter count in the primal space is not a meaningful measure of model complexity. We show the model with its maximum number of parameters possible and see that regularization to fewer parameters actually increases the performance for this model. The parameter counts are also not necessarily 1-to-1 comparable to those of a neural network model in terms of expressive power; nonetheless we include them for completeness.}. Details of each experiment are can be found in Section~\ref{sec:sweep_qubit_scaling}. \\
    }
  \label{tab:results}
\end{table}

\subsubsection{Accuracy over number of re-uploading layers}
\begin{figure}[hbt]
    \centering\includegraphics[width=0.45\textwidth]{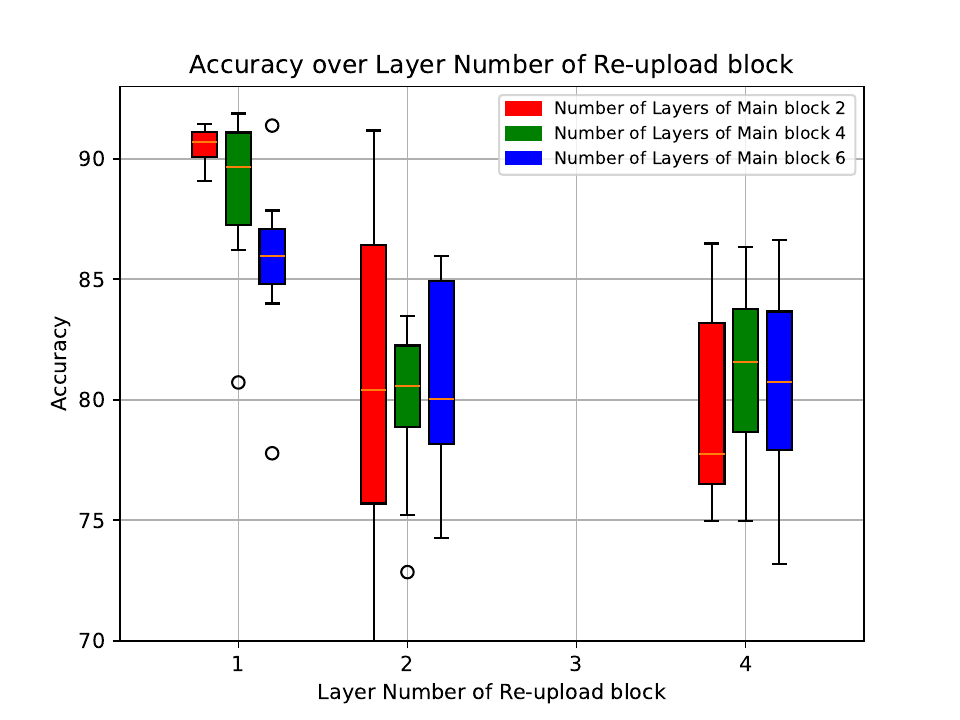}
    \caption{Accuracy as a function of number of re-uploading layers across number of main block layers decreases as the number of layers of re-uploading layers increases. This may be due to overparameterization and insufficient regularization associated with the increased circuit depth and the number of parameters. Hyperparameters chosen were: 
        \small
        $Q=14$,
        $E=1$,
        $R=4$,
        $B=16$,
        $L\in\{2,\,2.5,\,5\}\times10^{-3}$,
        $\gamma=1.0$,
        $\rho=0.0$,
        $S=8192$.
        }    
    \label{fig:acc_over_layer_num_reupload}
\end{figure}

The number of re-uploading layers determines the size of the PQC1 used in the repeat of Data re-uploading, as shown in Figure~\ref{fig:full_block_diagram}.
In Fig.~\ref{fig:acc_over_layer_num_reupload} shows the trend of accuracy over number of re-uploading layers across multiple number of main block layers. We tested 3 learning rates  and plotted the best results from each configuration, while keeping the re-upload number fixed.
The results of this investigation, which are summarized in Table~\ref{tab:config_hyper_cube_sweep_layer_num_reupload},  are inconclusive due to the decreasing accuracy as the number of layers increases. This may be due to the increased circuit depth and number of parameters from the greater number of re-uploading layers, which could lead to overparameterization and insufficient regularization. Further studies isolating the contributions of this effect would be valuable. 

\subsection{Ablation studies}
\subsubsection{Simulated quantum encoder}
We investigate the contribution of simulated quantum encoder to accuracy.
Table~\ref{tab:results_ablation_qie} shows the results of each configuration of neural-network-based encoder that replaces simulated quantum encoder. The data points are the best results of each configuration. The configurations are summarized in Table~\ref{tab:hyperparam_ablation_qie}.

Table~\ref{tab:results_ablation_qie} shows that simulated quantum encoder with two encoders yields the highest accuracy, followed by the neural network-based encoders and simulated quantum encoder with a single encoder. 
We observe that simulated quantum encoder with two encoders yields ($0.57$\%) higher accuracy than the highest accuracy (${92.13\%}$) of the neural network-based encoders. This indicates that simulated quantum encoder is effective. Additionally, simulated quantum encoder with a single encoder performs on par with the lowest accuracy of the neural network-based encoders. These observations suggest investigating the effect of increasing the number of simulated quantum encoder on accuracy. Consequently, this provides valuable insights for determining how to scale qubits for subsequent large-scale studies.

\begin{table}
    \centering
    \begin{tabular}{c c c} 
    \specialrule{.1em}{.05em}{.5em} 
    \multicolumn{3}{c}{\textbf{Simulated quantum encoder}} \\
    \specialrule{.05em}{.05em}{.5em} 
    \textbf{Number} & \textbf{Maximum } & \textbf{Number of } \\ 
    \textbf{of encoders}  & \textbf{Accuracy} & \textbf{parameters} \\ 
    \specialrule{.05em}{.05em}{.5em} 
     Single (1) & 91.45 & 493  \\ 
     Multi (2) & 92.70 & 886 \\ [0.5ex] 
    \specialrule{.1em}{.05em}{.5em} 
    \multicolumn{3}{c}{\textbf{Neural Network Encoder}} \\
    \specialrule{.05em}{.05em}{.5em} 
 \textbf{Hidden layers} & \textbf{Maximum } & \textbf{Number of }\\
 \textbf{/Dimensions} & \textbf{Accuracy} & \textbf{parameters} \\
    \specialrule{.05em}{.05em}{.5em} 
     0/0 & 91.92 &  10,866 \\
     1/48 & 92.08 & 37,698 \\
     1/96 & 91.46 & 75,282 \\
     1/144 & 92.13 & 112,866 \\
     1/192 & 91.93 & 150,450 \\ [1ex] 
    \specialrule{.05em}{.05em}{.5em} 
    \end{tabular}
    \caption{Results of ablation study replacing simulated quantum encoders with a neural network encoder. Hyperparameters chosen were:
        \small
        $Q=14$,
        $R=4$,
        $M=2$,
        $N=1$,
        $\gamma\in\{1.0,\,0.99,\,0.95\}$,
        $S=8192$,
        $B=16$,
        $L\in\{0.1,\,0.5,\,1,\,1.5,\,2,\,2.5,\,3,\,5,\,10,\,20\}\times10^{-3}$,
        $\rho\in\{0.0,\,1\times10^{-6},\,1\times10^{-5},\,1\times10^{-4},\,1\times10^{-3}\}$.
        }
    \label{tab:results_ablation_qie}
\end{table}

\subsubsection{Neural network-based classification head}
To evaluate the efficacy of our method, we performed an ablation study in which the method was replaced with a neural network-based classification Head. The configurations swept are summarized in Table~\ref{tab:hyperparam_ablation_nn}.
Table~\ref{tab:results} (bottom) represents the results for each configuration of the neural network-based classification head. The data points are the best results of each corresponding configuration. The results show that the highest accuracy achieved by the quantum fine-tuning method  (${92.70\%}$) is ($1.26$\%) higher than that of the best neural network-based classification head. Despite resource-limited, non-exhaustive exploration in a low-qubit regime, the comparison results are promising and motivate large-scale studies with extensive hyperparameter sweeps in a higher qubit regime, where we anticipate a more significant performance enhancement from the quantum fine-tuning method.

\subsubsection{Final linear layer}
Here, we study the contribution of final linear layer to accuracy.
Table~\ref{tab:results_ablation_final_lin_layer} shows that the effect of the final linear layer on accuracy. We observe that the configuration without the final linear layer yields slightly higher accuracy than the configuration with it. 
These observations offer architectural guidance for future large-scale studies.

\begin{table}
    \centering
    \begin{tabular}{c c}
\specialrule{.1em}{.05em}{.5em} 
        \textbf{Final Linear Layer} & \textbf{Maximum Accuracy} \\ [0.5ex] 
\specialrule{.05em}{.05em}{.05em} 
        Yes & 91.45 \\ 
        No & 91.70 \\ [0.5ex] 
\specialrule{.1em}{.05em}{.5em} 
    \end{tabular}
    \caption{Removing the final linear layer yields slightly higher accuracy. Hyperparameters chosen were: 
        \small
        $Q=14$,
        $E=1$,
        $R=4$,
        $M=2$,
        $N=1$,
        $B=16$,
        $\gamma=1.0$,
        $\rho=0.0$,
        $S=8192$,
        $L\in\{0.5,\,1,\,1.5,\,2,\,2.5,\,3,\,5\}\times10^{-3}$.
        }
    \label{tab:results_ablation_final_lin_layer}
\end{table}

\subsection{Gate noise studies}
\begin{figure}[hbt]
    \centering\includegraphics[width=0.9\linewidth]{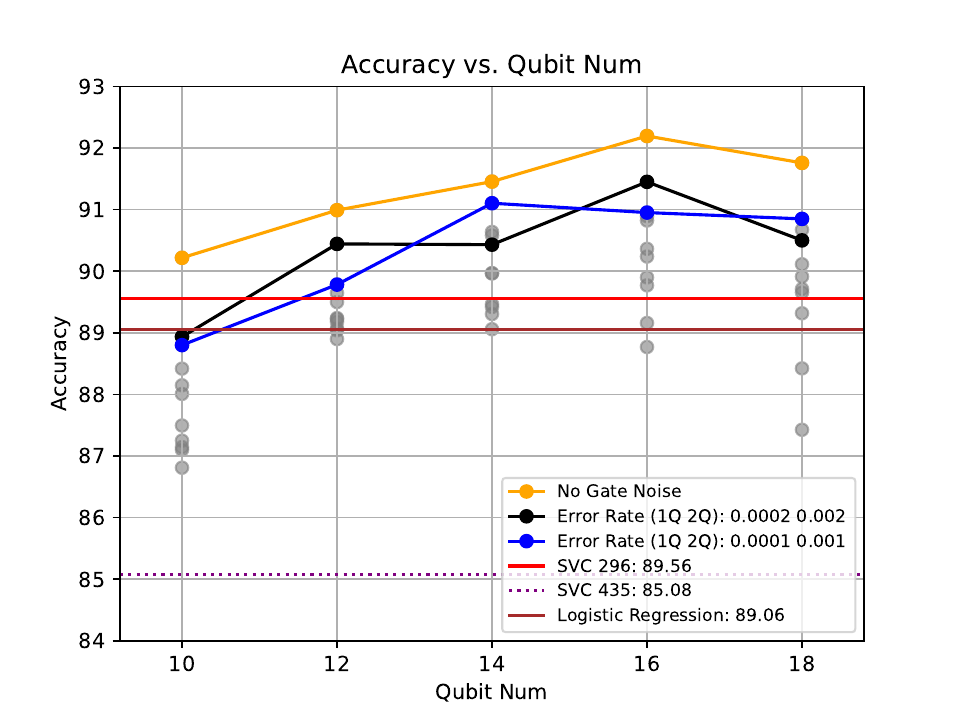}    
    \caption{Accuracy as a function of the number of qubits for two noise settings for the 1Q and 2Q gates. We observe lower absolute accuracy, but the same trend in accuracy with number of qubits. The chosen hyperparameters are:
        \small
        \small
        $Q=14$,
        $E=1$,
        $R=4$,
        $M=2$,
        $N=1$,
        $B=16$,
        $\gamma=1.0$,
        $\rho=0.0$,
        $S=8192$,
        $L\in\{1,\,1.5,\,2,\,2.5,\,3,\,5\}\times10^{-3}$.
        }    
    \label{fig:max_acc_gate_noise}
\end{figure}

Figure~\ref{fig:max_acc_gate_noise} illustrates accuracy as a function of qubit number with gate noise for the single-encoder architecture. Consistent with the qubit scaling experiments, the ansatz configurations (re-upload number, number of main block layers, number of re-uploading layers) are held constant. The training hyperparameters are detailed in Table~\ref{tab:config_gn_sweep}. Considering practical availability of computational resources (GPUs), we restricted our hyperparameter sweep to the learning rate only. Despite the presence of gate noise, the same upward trend in accuracy with qubit number is observed. The noisy plots follow the overall contour of the plot without gate noise, albeit with lower absolute accuracy. As indicated earlier, by optimizing the hyperparameters at each qubit number---rather than keeping them fixed for the extent of the experiment--there is margin for all of these values to increase in accuracy.

\subsection{Empirical upper bound}
To put into perspective the performance enhancement provided by quantum fine-tuning an LLM with frozen body, we consider an empirical upper bound. For this, we performed a separate experiment in which we performed end-to-end fine-tuning of both the LLM body and a differentiable classical classification head. We consider this an upper bound because we allow the body to be updated as well, whereas in our experiments, we freeze the body. A summary of the hyperparameter sweeps and the results is presented in Table~\ref{tab:emp_upper_boud}. We obtained an upper bound on the achievable accuracy of about $93.62\%$, showing that the hybrid quantum/classical model comes already close in performance, even with frozen weights in the body of the LLM.

\section{Conclusion}
We introduced a hybrid quantum-classical architecture for language model fine-tuning, including results from a screening study to explore the main effects of sweeping over a finite set of variables. We study the scaling in task accuracy with both circuit depth and number of qubits, identifying a promising trend of accuracy increasing with number of qubits, while depth has a width-dependent optimal setting apparently balancing expressive power and trainability.
One direction for future work is to perform full fine-tuning by unfreezing the weights of the base language model. 
Another avenue is a design-of-experiment based approach to optimize hyperparameters beyond our study on supercomputers, such as ORNL Frontier \cite{noauthor_frontier_nodate}, extending our results for higher qubit counts across a broader range of circuit depths. Finally, we plan to validate our findings by performing inference and training on physical quantum hardware.

\cleardoublepage
\newpage
\section{Supplementary Information}

\subsection{SST2 Dataset}
We use the Stanford Sentiment Treebank (SST2), specifically, the version trimmed for SetFit\footnote{\url{https://huggingface.co/datasets/SetFit/sst2}}. The dataset contains $N=9613$ total sentences labeled as either positive ($+$) or negative ($-$) connotation in the sentiment of their content. We first take $256$ samples from each label $\{+, -\}$ as usable during training, and further split this as $85\%$ for a ``training'' set and $15\%$ for a ``validation'' set. The remaining ``test'' split of the data set contains $N-2\times 256 = 9101$ total samples from both labels.

\label{sec:sst2_for_setfit} 

\subsection{Energy consumption estimate}

\begin{table}[hbt]
    \centering
    \begin{tabular}{c c}
\specialrule{.1em}{.05em}{.05em} 
        QPU energy consumption ($P_{qpu}$) & 300W/qubit \\         
        1Q gate time ($T_{sq}$) & $10^{-4}$s \\
        2Q gate time ($T_{tq}$) & $10^{-5}$s \\ 
        Number of shots ($S$) & 8,000 \\
        H100 GPU energy consumption ($P_{gpu}$) & 700W/GPU \\ 
        H100 GPU FP64 speed ($F_{gpu}$) & $3.4 \times 10^{13}$ FLOPS \\ [1ex] 
\specialrule{.1em}{.05em}{.05em} 
    \end{tabular}
    \caption{Specifications used for estimating energy consumption for both QPU and GPU.}
    \label{tab:energy_consumption_assumption}
\end{table}

$E_{qpu}$ and $E_{gpu}$ are the estimated energy consumption for QPU and GPU, respectively.

$$ E_{qpu} = N \cdot ( SQ \cdot T_{sq} + TQ \cdot T_{tq}) \cdot S \cdot \frac{P_{qpu}}{1000} kJ $$ 
$$ E_{gpu} = 2^{N} \cdot \frac{(SQ \cdot 2^{2} + TQ \cdot 2^{3})}{F_{gpu}} \cdot \frac{P_{qpu}}{1000} kJ $$ \\
where $N$ is the number of qubits, $SQ$ is single qubit gate count, and $TQ$ is 2-qubit gate count.

\label{sec:energy_consumption_cmp} 

We assume that the GPU communication overhead is nonexistent even when the statevector is too large to fit on a single device and needs to be distributed across multiple devices. We note that as a result, the number of GPUs required cancels out in our computations, as it contributes the same multiplicative factor to both the total FLOPS in the denominator and the total power in the numerator. Overall, this is optimistic in favor of the GPU, so produces a more conservative estimate of the crossover point.

\subsection{Loss}

\begin{figure}[hbt]
    \centering\includegraphics[width=0.75\linewidth]{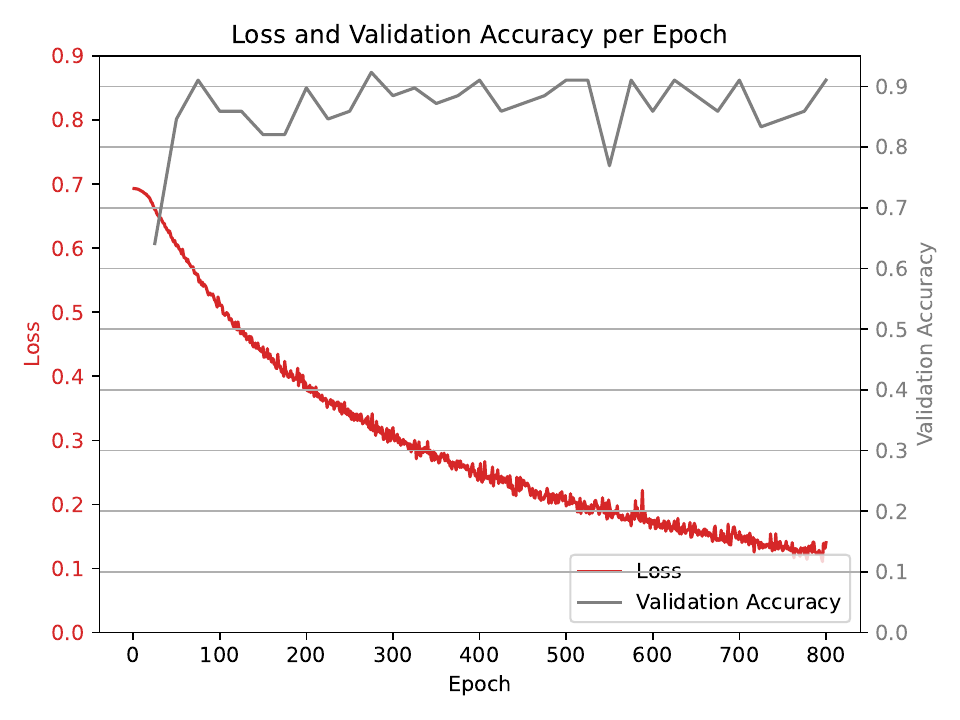}
    \caption{Loss and validation accuracy over epoch: shown is the training loss and validation accuracy of the configuration multi-encoder sQE 18Q over 800 epochs.}   
    \label{fig:loss_plot}
\end{figure}

\label{sec:training_loss} 

\subsection{Hyperparameter configurations}
\label{sec:sweep_qubit_scaling}

\begin{table}[hbt]
    \centering
    \begin{tabular}{c c}
\specialrule{.1em}{.05em}{.05em} 
        Qubits & 10, 12, 14, 16, 18 \\         
        re-upload number & 4 \\ 
        number of main block layers & 2  \\
        number of re-uploading layers & 1 \\         
        Connectivity & 1 \\ 
        Batch Size & 16 \\ 
        Number of shots & 8192 \\ 
        Final Linear Layer & Yes \\ 
        Learning rate & $\{1, 1.5, 2, 2.5, 3, 5\} / 10^{3}$ \\     
        Learning rate decay & 1.0 \\ 
        Weight decay & 0.0 \\ [1ex] 
\specialrule{.1em}{.05em}{.05em} 
    \end{tabular}
    \caption{Configurations of hyperparameter sweep for single simulated quantum encoder.}
    \label{tab:config_se_sweep}
\end{table}

\begin{table}[hbt]
    \centering
    \begin{tabular}{cc}
\specialrule{.1em}{.05em}{.05em} 
        Qubits & 10, 12, 14, 16, 18 \\ 
        Number of Encoders & 2 \\         
        re-upload number & 4 \\ 
        number of main block layers & 2  \\ 
        number of re-uploading layers & 1 \\         
        Connectivity & 1 \\ 
        Batch Size & 16 \\ 
        Number of shots & 8192 \\ 
        Final Linear Layer & Yes \\ 
        Learning rate & $\{1, 1.5, 2.5, 3, 5\}/ 10^{3}$ \\         
        Learning rate decay & 1.0 \\ 
        Weight decay & 0.0 \\ [1ex] 
\specialrule{.1em}{.05em}{.05em} 
    \end{tabular}
    \caption{Configurations of hyperparameter sweep for multi simulated quantum encoder.}
    \label{tab:config_me_sweep}
\end{table}

\begin{table}[hbt]
    \centering
    \begin{tabular}{cc}
\specialrule{.1em}{.05em}{.05em} 
        C (Regularization) &  0.001, 0.005, 0.01, 0.05, 0.1, \ldots, 100.0 \\ 
        Kernel & RBF \\ 
        Gamma (kernel coeff) &  scale, auto, 0.01, 0.1, 0.5, 1.0 \\ [1ex] 
\specialrule{.1em}{.05em}{.05em} 
    \end{tabular}
    \caption{Configurations of SVC.}
    \label{tab:config_svc}
\end{table}

\begin{table}[hbt]
    \centering
    \begin{tabular}{cc}
\specialrule{.1em}{.05em}{.05em} 
        Penalty &  0.3, 0.35, 0.4, \ldots, 9.90 \\
        C (Regularization) & l2, None \\
        Solver &  lbfgs, \\
        Class Weight & balanced, None, \\
        Max Iteration & 50, 100, 200 \\ [1ex] 
\specialrule{.1em}{.05em}{.05em} 
    \end{tabular}
    \caption{Configurations of Logistic Regression.}
    \label{tab:config_lrg}
\end{table}

\begin{table}[hbt]
    \centering
    \begin{tabular}{cc}
\specialrule{.1em}{.05em}{.05em} 
        Qubits & 14 \\  
        re-upload number & 4, 6, 8 \\
        number of main block layers & 2, 3, 4, 5, 6, 7, 8  \\
        number of re-uploading layers & 1 \\        
        Connectivity & 1 \\
        Batch Size & 16 \\ 
        Number of shots & 8192 \\
        Final Linear Layer & Yes \\
        Learning rate & $\{2, 2.5, 5\} / 10^{3}$ \\
        Learning rate decay & 1.0 \\ 
        Weight decay & 0.0 \\ [1ex]
\specialrule{.1em}{.05em}{.05em} 
    \end{tabular}
    \caption{Configurations of hyperparameter sweep for accuracy over re-upload number.}
    \label{tab:config_hyper_cube_sweep_reupload_num}
\end{table}

\begin{table}[hbt]
    \centering
    \begin{tabular}{cc}
\specialrule{.1em}{.05em}{.05em} 
        Qubits & 14 \\  
        re-upload number & 4 \\
        number of main block layers & 2, 4, 6 \\
        number of re-uploading layers & 1, 2, 4 \\
        Connectivity & 1 \\
        Batch Size & 16 \\ 
        Number of shots & 8192 \\ 
        Final Linear Layer & Yes \\
        Learning rate & $\{2, 2.5, 5\} / 10^{3}$ \\
        Learning rate decay & 1.0 \\ 
        Weight decay & 0.0 \\ [1ex]
\specialrule{.1em}{.05em}{.05em} 
    \end{tabular}
    \caption{Configurations of hyperparameter sweep for accuracy over number of re-uploading layers.}
    \label{tab:config_hyper_cube_sweep_layer_num_reupload}
\end{table}

\begin{table}[hbt]
    \centering
    \begin{tabular}{cc}
\specialrule{.1em}{.05em}{.05em} 
        Qubits & 14 \\  
        re-upload number & 4 \\ 
        number of main block layers & 2  \\
        number of re-uploading layers & 1 \\         
        Connectivity & 1 \\ 
        Batch Size & 16 \\ 
        Number of shots & 8192 \\ 
        Final Linear Layer & Yes, No \\ 
        Learning rate & $\{0.5, 1, 1.5, 2, 2.5, 3, 5\} / 10^{3}$ \\     
        Learning rate decay & 1.0 \\ 
        Weight decay & 0.0 \\ [1ex] 
\specialrule{.1em}{.05em}{.05em} 
    \end{tabular}
    \caption{Configurations of hyperparameter sweep for ablation study for final linear layer.}
    \label{tab:config_fin_lin_layer}
\end{table}

\begin{table}[hbt]
    \centering
    \begin{tabular}{cc}
\specialrule{.1em}{.05em}{.05em} 
        Hidden Layers & 0, 1\\
        Hidden Dimensions & 48, 96, 144, 192 (0 for Hidden Layer 0)  \\
        Batch Norm  & Yes or No \\
        Qubits & 14 \\  
        re-upload number & 4 \\ 
        number of main block layers & 2  \\
        number of re-uploading layers & 1 \\
        Connectivity & 1 \\ 
        Batch Size & 16 \\ 
        Number of shots & 8192 \\ 
        Final Linear Layer & Yes \\ 
        Learning rate & $\{0.1, 0.5, 1, 1.5, 2, 2.5, 3, 5, 10, 20\} / 10^{3}$ \\        
        Learning rate decay & 1.0, 0.99, 0.95 \\
        Weight decay & 0.0, 1e-6, 1e-5, 1e-4, 1e-3 \\ [1ex] 
\specialrule{.1em}{.05em}{.05em} 
    \end{tabular}
    \caption{Configurations of hyperparameter sweep for ablation study for simulated quantum encoder.}
    \label{tab:hyperparam_ablation_qie}
\end{table}

\begin{table}[hbt]
    \centering
    \begin{tabular}{cc}
\specialrule{.1em}{.05em}{.05em} 
        Hidden Layers & 0, 1\\
        Hidden Dimensions & 48, 96, 144, 192 (0 for Hidden Layer 0) \\
        Batch Norm  & Yes or No \\ 
        Learning rate & $\{0.1, 0.5, 1, 1.5, 2, 2.5, 3, 5, 10, 20\} / 10^{3}$ \\        
        Learning rate decay & 1.0, 0.99, 0.95 \\
        Weight decay & 0.0, 1e-6, 1e-5, 1e-4, 1e-3 \\ [1ex] 
\specialrule{.1em}{.05em}{.05em} 
    \end{tabular}
    \caption{Configurations of hyperparameter sweep for ablation study for neural network head.}
    \label{tab:hyperparam_ablation_nn}
\end{table}

\begin{table}[hbt]
    \centering
    \begin{tabular}{c c}
\specialrule{.1em}{.05em}{.05em} 
        Qubits & 10, 12, 14, 16, 18 \\  
        Error Rate (1Q 2Q) & (0.0001 0.001), (0.0002 0.002) \\
        re-upload number & 4 \\ 
        number of main block layers & 2  \\
        number of re-uploading layers & 1 \\ 
        Connectivity & 1 \\ 
        Batch Size & 16 \\ 
        Number of shots & 8192 \\ 
        Final Linear Layer & Yes \\ 
        Learning rate & $\{1, 1.5, 2, 2.5, 3, 5\} / 10^{3}$ \\     
        Learning rate decay & 1.0 \\ 
        Weight decay & 0.0 \\ [1ex] 
\specialrule{.1em}{.05em}{.05em} 
    \end{tabular}
    \caption{Configurations of hyperparameter sweep for gate noise studies.}
    \label{tab:config_gn_sweep}
\end{table}

\begin{table}[hbt]
    \centering
    \begin{tabular}{cc}
\specialrule{.1em}{.05em}{.05em} 
        Hidden Layers & 0, 1\\
        Hidden Dimensions & 48, 96, 144, 192 (0 for Hidden Layer 0) \\
        Batch Size & 16 \\ 
        Batch Norm  & Yes, No \\ 
        Learning rate & $\{0.01, 0.02, 0.05, 0.1, 0.5, 1.0, 2.0, 5.0\} / 10^{3}$ \\
        Learning rate decay & 1.0, 0.99, 0.95 \\
        Weight decay & 0.0, 1e-6, 1e-5, 1e-4, 1e-3 \\  
        Max. Accuracy & 93.62 \\ [1ex]
\specialrule{.1em}{.05em}{.05em} 
    \end{tabular}
    \caption{Configurations of hyperparameter sweep for obtaining empirical upper bound through end-to-end training of both the LLM and a differentiable classification head.}
    \label{tab:emp_upper_boud}
\end{table}

\end{document}